# Transmission Line Inspires A New Distributed Algorithm to Solve the Linear System of Circuit

Fei Wei, Huazhong Yang, *Senior Member, IEEE*

*Abstract*—Transmission line, or wire, is always troublesome to integrated circuits designers, but it could be helpful to parallel computing researchers. This paper proposes the Virtual Transmission Method (VTM), which is a new distributed and stationary iterative algorithm to solve the linear system extracted from circuit. It tears the circuit by virtual transmission lines to achieve distributed computing. For the symmetric positive definite (SPD) linear system, VTM is proved to be convergent. For the unsymmetrical linear system, numerical experiments show that VTM is possible to achieve better convergence property than the traditional stationary algorithms. VTM could be accelerated by some preconditioning techniques, and the convergence speed of VTM is fast when its preconditioner is properly chosen.

## I. INTRODUCTION

TRANSMISSION line, or interconnect, or wire, is one of the main determinants of chip performance, because of the long interconnect latency and large power dissipation [37]. However, the circuit could not work without transmission lines, because the transmission line plays an irreplaceable role to make the distributed circuit stable and scalable [13]. In this paper, transmission lines are used to assist the distributed computation of sparse linear system of circuit.

The basic problem of the distributed circuit simulation is to distributedly solve the sparse linear system extracted from circuit [34, 41, 46]. Generally speaking, the linear system of circuit is unsymmetrical, because of the existence of controlled sources. Restricted to the resistor-capacitor (RC) network, the linear system would be symmetrical [39].

Many numerical algorithms are employed to solve the linear system from circuit. KLU is an optimized LU factorization algorithm for circuit, and it is broadly used in commercial and non-commercial circuit simulators [49]. KLU is sequential. SuperLU is an outstanding distributed algorithm to solve the general unsymmetrical sparse linear system on a large number of processors, and it has been widely adopted for circuit simulation [15]. Xyce is a parallel circuit simulator designed for supercomputers, and it uses the Block Jacobi preconditioned GMRES method [50, 4, 3]. Schur complement method is a non-overlapping domain decomposition method making use of the master-slave model. It is simple to be comprehended and implemented, and was frequently used [29, 65, 20]. [7] presents a parallel direct/iterative mixed approach, based on the Schur complement method and preconditioned GMRES method. [41] adopted the preconditioned overlapping domain decomposition method, which is supported by PETSc, a distributed scientific computing package [1]. The highlight of [41] is that it improves the matrix property by exploring the characters of transistor device models.

The symmetric linear system of circuit is mainly extracted from the RC networks. Solving RC network is the key for the power grid analysis and thermal analysis [30, 26]. Meanwhile, the sparse linear system generated by the finite element method from elliptic partial differential equations (PDE) is equivalent to a resistor network.

The simulation of power grid has been intensively researched in the past 10 years, and a variety of numerical algorithms had been employed to solve the RC network, such as stationary iterative methods [67], Krylov-subspace iterative method [12], algebraic multigrid (AMG) [31, 69], coupled iterative/direct method [32] and FFT method [48]. [42] proposed a stochastic method, and this method was further expanded as a preconditioning method for the diagonal dominant matrix [43].

To compute the RC network on the parallel computing platform, [68] makes use of the domain decomposition method (DDM). [11] adopts a distributed matrix inversion algorithm. The potential of GPU to solve the RC network was illustrated in [6, 16, 17, 48].

VTM is a new distributed and stationary iterative algorithm to solve the sparse linear systems [59]. VTM is similar to Block Jacobi or Overlapped Block Jacobi method [56, 66].

VTM is inspired by the behavior of transmission line [40, 13]. It inserts virtual transmission lines into the circuit to achieve distributed computing [58, 59]. The idea of VTM is derived from the pseudo transient method [57]. Pseudo transient method inserts pseudo capacitors or inductors into the circuit to approve the convergence property of DC analysis. Similarly, VTM inserts pseudo transmission lines into the circuit to approve the convergence property of distributed simulation. Inserting virtual or pseudo electrical element to assist the circuit simulation is a common approach, and the advantage is that the physical property of circuit could be utilized during numerical computation.

VTM draws a parallel between distributed computing of linear system and distributed simulation of linear circuit, and it interprets the relationship between numerical analysis and circuit analysis. To solve a sparse linear system in parallel is equalized to simulate a linear circuit in parallel, and vice versa, we might comprehend the distributed numerical algorithm from the viewpoint of circuit theory and microwave network.

To partition a circuit, node tearing and branch tearing are two important methods [64, 47]. This paper proposes the third way, which is called wire tearing. Tearing the circuit by wires is not a new idea, since all the subcircuits are connected by wires, but we first apply this concept into the distributed numerical algorithms [58, 59]. Wire tearing can be considered as the insertion of transmission line to connect the torn nodes [61]. The advantage of wire tearing over the branch tearing and node tearing is that it does not bring in any extra energy, because all the wires are passive [27].

As a distributed numerical algorithm, VTM is similar to overlapped Block Jacobi or Block SOR method [2, 45]. We have proved that VTM is convergent to solve arbitrarily-large SPD linear system on arbitrary number of processors [59, 62].

VTM could also be considered as a new type of algebraic domain decomposition method [55]. VTM is similar to the additive Schwarz method with Robin condition, i.e. Schwarz-Robin method [35, 21, 22, 36]. Additive Schwarz method is mainly used to solve the sparse linear system from elliptic partial differential equations (PDE), and VTM is used to solve the general sparse linear system. The partitioning method for VTM, i.e. wire tearing, is different from the traditional partitioning methods for algebraic additive Schwarz method [22, 36]. Furthermore, the preconditioning technique for VTM is more flexible than Optimized Schwarz Method [21].

This paper is organized as follows. Section 2 presents a brief introduction to the linear system of circuit. Section 3 introduces the mathematical description of transmission line. Section 4 proposes the physical background of VTM. Section 5 describes how to partition the circuit by wire tearing. Section 6 details the algorithm of VTM. Section 7 presents the convergence theory. Section 8 provides a simple example. Section 9 discusses the preconditioning techniques for VTM. Numerical experiments are shown in Section 10. We conclude this work in Section 11. In the appendix, we present a proof for the convergence theory.

## II. Linear System of Circuit

In virtue of the nodal analysis, one circuit could be described by a sparse linear system [39].

$$\mathbf{Ax} = \mathbf{b} \qquad (1)$$

Here **x** represents the nodes' voltages, and **b** is the current sources flowing into the nodes. **A** is the coefficient matrix, which might be symmetrical, or unsymmetrical, depending on the property of circuit.

If the circuit contains no voltage sources but only current sources, we call it current-driven circuit. To construct the linear system of current-driven circuit, we only need to use the nodal analysis technique, and do not need the modified nodal analysis technique [39]. If there were voltage sources in the circuit, they might be equalized into current sources, according to the Norton equivalent theory [18].

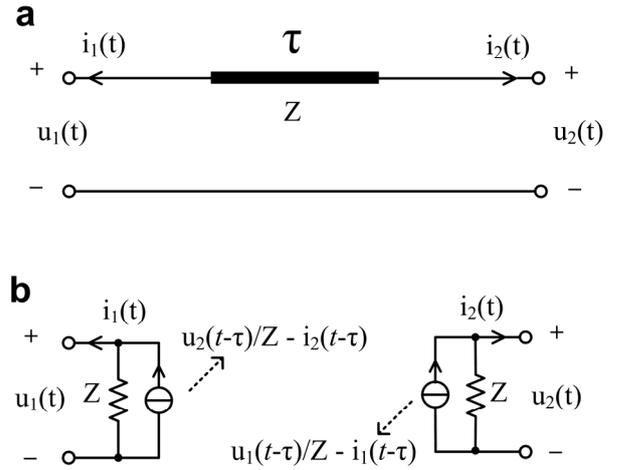

Fig. 1. Transmission line. (A) The circuit diagram of the transmission line. (B) The equivalent circuit of the lossless transmission line.

**Lemma 1**: For a current-driven linear resistor network, its coefficient matrix, **A**, is weakly diagonally dominant, and thus symmetrical-non-negative-definite (SNND). Further, if there is at least one resistor connecting to the ground, **A** is strictly diagonally dominant, and thus symmetric-positive-definite (SPD).

The physical insight of this lemma is that the energy of a resistor network is always non-negative.

## III. Transmission Line

Transmission line is an important element in electrical and microwave engineering. It is also called cable, wire or interconnect in different contexts.

The circuit diagram of the transmission line is illustrated in Fig. 1. The analytical description of the lossless or ideal transmission line is the wave equation [40, 13].

$$\frac{\partial^2 u(x,t)}{\partial^2 x} = LC \frac{\partial^2 u(x,t)}{\partial^2 t} \qquad (2)$$

Here $L$ is the inductance per unit length, $C$ is the capacitance per unit length.

The time domain mathematical description of the lossless transmission line is called Transmission Delay Equations, as shown in (3) [40, 8, 59].

$$\begin{aligned} u_1(t) + Z \cdot i_1(t) &= u_2(t-\tau) - Z \cdot i_2(t-\tau) \\ u_2(t) + Z \cdot i_2(t) &= u_1(t-\tau) - Z \cdot i_1(t-\tau) \end{aligned} \qquad (3)$$

Here $u_1(t)$ and $u_2(t)$ represent the port voltages. $i_1(t)$ and $i_2(t)$ represent port inflow currents. $t$ is the time variable. $\tau$ is the propagation delay. Z is the characteristic impedance, which is positive.

$$Z = \sqrt{L/C}$$

The propagation delay of the transmission line is:

$$\tau = l\sqrt{LC}$$

Where $l$ is the length of the transmission line.

The iterative formula of Transmission Delay Equations is similar to the Robin boundary condition in Partial Differential Equations and Schwarz-Robin method [35, 21]. There are two main differences: First, the normal derivative in the robin condition is replaced by a current; Second, (3) expresses the time delay explicitly.

For a number of coupled transmission lines, (3) could be re-expressed in the matrix-vector form:

$$\begin{aligned}\mathbf{u}_1(t) + \mathbf{Z} \cdot \mathbf{i}_1(t) &= \mathbf{u}_2(t-\tau) - \mathbf{Z} \cdot \mathbf{i}_2(t-\tau) \\ \mathbf{u}_2(t) + \mathbf{Z} \cdot \mathbf{i}_2(t) &= \mathbf{u}_1(t-\tau) - \mathbf{Z} \cdot \mathbf{i}_1(t-\tau)\end{aligned} \quad (4)$$

Here $\mathbf{Z}$ is the characteristic impedance matrix of the transmission lines. If there is no coupling among transmission lines, $\mathbf{Z}$ is diagonal; if there is coupling, $\mathbf{Z}$ is a sparse matrix.

(4) could be re-expressed as (5):

$$\begin{aligned}\mathbf{i}_1(t) + \mathbf{W} \cdot \mathbf{u}_1(t) &= -\mathbf{i}_2(t-\tau) + \mathbf{W} \cdot \mathbf{u}_2(t-\tau) \\ \mathbf{i}_2(t) + \mathbf{W} \cdot \mathbf{u}_2(t) &= -\mathbf{i}_1(t-\tau) + \mathbf{W} \cdot \mathbf{u}_1(t-\tau)\end{aligned} \quad (5)$$

Here $\mathbf{W}$ is the characteristic admittance matrix of the wires.

$$\mathbf{W} = \mathbf{Z}^{-1}$$

Associated with the principle of parallel computer, we come to realize that, there is conceptual similarity between the distributed computer and the distributed circuit, as shown in Fig. 2 [61].

First, different circuits are connected by wires, which is similar to the situation where different processors are connected by the digital data link.

Second, the wire has transmission delay, and there is also communication delay between the processors.

As the result, if we assimilate each circuit to a processor, and consider the wire as the data link between processors, then the distributed circuit would be kind of a distributed computer.

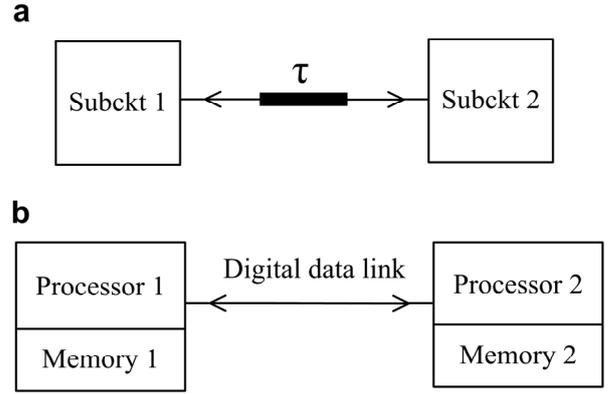

Fig. 2. Similarity between the distribute circuit and distributed computer. (**a**) Distributed circuit, two subcircuits are connected by a wire. (**b**) Distributed computer, two processors are connected by a digital data link.

## IV.  Physical Background

In this section, we present an observation from circuit and microwave network. This lemma presents the stability property of the resistor network with wires. Fig. 3 presents an illustration for this lemma. Lemma 2 could be validated in SPICE [38, 44].

**Lemma 2**: A linear resistor network with wires inside never oscillates unendingly, and it will finally settle down to the steady state, which is the steady state of the resistor network eliminating all the wires.

The physical insight of this lemma is that, the energy of a resistor network is limited. Divergence means infinite energy, which is impossible for the resistor network.

Lemma 2 is also valid for the resistor-capacitor network. But for the general circuit with wires, it is not always valid. A distributed circuit with wires inside might go unconvergent. The stability is depending on the property of circuit and the characteristic impedance of the wires.

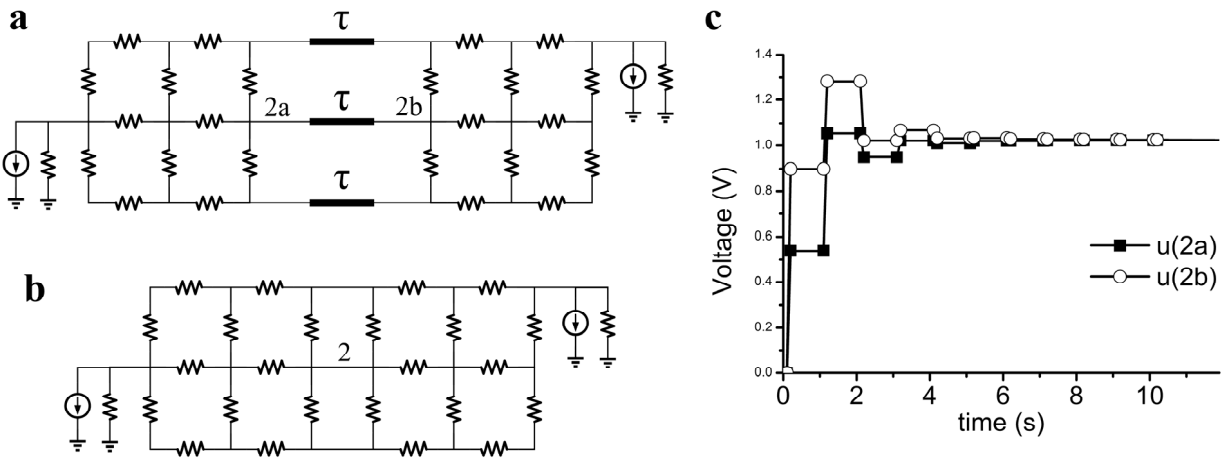

Fig. 3.  Illustration of the stability theory of resistor network with wires. (**a**) The resistor network with wires inside. (**b**) The resistor network without wires. (**c**) The voltage of node 2a and node 2b in (a) woud converge to the steady state, which is the voltage of node 2 in (b).

## V. Wire Tearing

To solve the sparse linear system of a circuit, originally there is no transmission line in it, but we might figure out some way to insert transmission lines into the circuit by splitting the nodes. This partitioning technique is called wire tearing.

By means of wire tearing, the original circuit is turned into a distributed circuit. It is partitioned into a number of separate sub-circuits by virtual transmission line. Later, we locate each sub-circuit into one processor, and use the digital data link to imitate the behavior of the transmission line. As the result, the distributed circuit is simulated in a distributed way.

According to Lemma 2, the resistor network with wires would finally go to the steady state, which is exactly the steady state of the resistor network without wires. This lemma explains the reason that VTM could always be convergent to solve the resistor network. For the sparse linear system of a general circuit, VTM might be unconvergent.

There are 4 steps to perform wire tearing for the circuit $G$. $\Gamma$ denotes the set of all the nodes in $G$.

Step 1. Set the splitting interface $\Gamma_{interface} \subseteq \Gamma$. V is called interfacial node if $V \in \Gamma_{interface}$; otherwise, V is called inner node.

Step 2. Split each interfacial node into a pair of nodes, which are called twin nodes.

Step 3. Split the resistors and current sources connected to each interfacial node.

Step 4. Connect each pair of twin nodes by a length of lossless wire. Add inflow currents to the twin nodes.

**Example 1**. Fig. 4 illustrates an example, in which a simple resistor network is split into two sub-circuits.

First we define the set of inner nodes in Sub-circuit 1 as $\Gamma_{1,inner}$, the set of inner nodes in Sub-circuit 2 as $\Gamma_{2,inner}$.

By reordering the rows, the sparse linear system of circuit (1) could be re-formatted as below by row reordering:

$$\begin{bmatrix} \mathbf{C} & \mathbf{E_1} & \mathbf{E_2} \\ \mathbf{F_1} & \mathbf{D_1} & 0 \\ \mathbf{F_2} & 0 & \mathbf{D_2} \end{bmatrix} \begin{bmatrix} \mathbf{u} \\ \mathbf{y_1} \\ \mathbf{y_2} \end{bmatrix} = \begin{bmatrix} \mathbf{f} \\ \mathbf{g_1} \\ \mathbf{g_2} \end{bmatrix} \quad (6)$$

Here $\mathbf{u}$ is the voltage vector of $\Gamma_{interface}$. $\mathbf{y_1}$ and $\mathbf{y_2}$ are the voltage vector of $\Gamma_{1,inner}$ and $\Gamma_{2,inner}$, respectively.

After wire tearing, the interfacial nodes are split, and $\Gamma_{interface}$ is split into two sets, the twin nodes in Sub-circuit 1, $\Gamma_{1,twin}$, and the twin nodes in Sub-circuit 2, $\Gamma_{2,twin}$. The system of sub-circuit 1 is:

$$\begin{bmatrix} \mathbf{C_1} & \mathbf{E_1} \\ \mathbf{F_1} & \mathbf{D_1} \end{bmatrix} \begin{bmatrix} \mathbf{u_1} \\ \mathbf{y_1} \end{bmatrix} = \begin{bmatrix} \mathbf{f_1} \\ \mathbf{g_1} \end{bmatrix} + \begin{bmatrix} \mathbf{i_1} \\ 0 \end{bmatrix} \quad (7)$$

$\mathbf{u_1}$ represents the voltage vector of $\Gamma_{1,twin}$. $\mathbf{i_1}$ represents the currents flowing into $\Gamma_{1,twin}$.

After wire tearing, the system of sub-circuit 2 is:

$$\begin{bmatrix} \mathbf{C_2} & \mathbf{E_2} \\ \mathbf{F_2} & \mathbf{D_2} \end{bmatrix} \begin{bmatrix} \mathbf{u_2} \\ \mathbf{y_2} \end{bmatrix} = \begin{bmatrix} \mathbf{f_2} \\ \mathbf{g_2} \end{bmatrix} + \begin{bmatrix} \mathbf{i_2} \\ 0 \end{bmatrix} \quad (8)$$

Here $\mathbf{u_2}$ represents the voltages vector of $\Gamma_{2,twin}$. $\mathbf{i_2}$ represents the currents flowing into each node of $\Gamma_{2,twin}$.

Also we have,

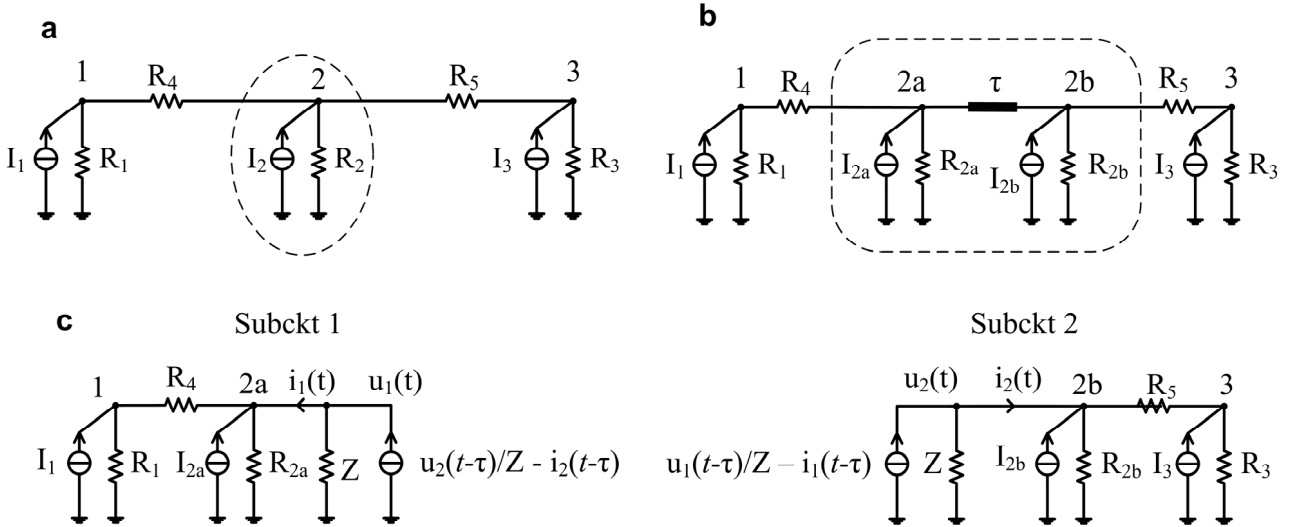

Fig. 4. Illustration of the wire tearing. (**a**) The original resistor network, and node 2 is set to be the interfacial node. (**b**) Split node 2 into a pair of twin nodes 2a and 2b, and insert a length of transmission line between them. $1/R_2 = 1/R_{2a} + 1/R_{2b}$. $I_2 = I_{2a} + I_{2b}$. (**c**) Replace the virtual transmission line with its equivalent circuit, and the original resistor network is partitioned into two sub-networks.

$$\mathbf{C_1} + \mathbf{C_2} = \mathbf{C}$$
$$\mathbf{f_1} + \mathbf{f_2} = \mathbf{f}$$

$\mathbf{u_1}$, $\mathbf{i_1}$, $\mathbf{u_2}$, $\mathbf{i_2}$ are called boundary variables, and they satisfy the Transmission Delay Equations (5).

$$\mathbf{i_1}(t) + \mathbf{W} \cdot \mathbf{u_1}(t) = -\mathbf{i_2}(t-\tau) + \mathbf{W} \cdot \mathbf{u_2}(t-\tau)$$
$$\mathbf{i_2}(t) + \mathbf{W} \cdot \mathbf{u_2}(t) = -\mathbf{i_1}(t-\tau) + \mathbf{W} \cdot \mathbf{u_1}(t-\tau)$$

If we set the boundary condition as below:

$$\mathbf{u_1}(t) = \mathbf{u_2}(t) = \mathbf{u}(t)$$
$$\mathbf{i_1}(t) = -\mathbf{i_2}(t) = \mathbf{i}(t)$$

Then (6) could be re-obtained from (7) and (8).

The above splitting technique is called level-one wire tearing, since the interfacial node is split once. If the interfacial node was split for more than one times, this splitting technique is called multilevel wire tearing [59].

## VI. VIRTUAL TRANSMISSION METHOD

After partitioning of the circuit, there are 3 steps to perform the parallel computing by Virtual Transmission Method.
  Step 1. Replace the lossless transmission line with its equivalent circuit.
  Step 2. Load each sub-circuit into a processor.
  Step 3. Use the data link to transfer the previous interfacial variables from one processor to another by message passing, and perform the distributed computing.

**Example 2**. Continuing with Example 1, the analytic expression of sub-circuit 1 is obtained by merging (5) and (7):

$$\begin{bmatrix} \mathbf{C_1} & \mathbf{E_1} \\ \mathbf{F_1} & \mathbf{D_1} \end{bmatrix} \begin{bmatrix} \mathbf{u_1}(t) \\ \mathbf{y_1}(t) \end{bmatrix} = \begin{bmatrix} \mathbf{f_1} \\ \mathbf{g_1} \end{bmatrix} + \begin{bmatrix} \mathbf{i_1}(t) \\ 0 \end{bmatrix} \quad (9)$$
$$\mathbf{i_1}(t) + \mathbf{W} \cdot \mathbf{u_1}(t) = -\mathbf{i_2}(t-\tau) + \mathbf{W} \cdot \mathbf{u_2}(t-\tau)$$

Eliminate $\mathbf{i_1}(t)$, and we obtain:

$$\begin{bmatrix} \mathbf{C_1} + \mathbf{W} & \mathbf{E_1} \\ \mathbf{F_1} & \mathbf{D_1} \end{bmatrix} \begin{bmatrix} \mathbf{u_1}(t) \\ \mathbf{y_1}(t) \end{bmatrix} = \begin{bmatrix} \mathbf{f_1} + \mathbf{W} \cdot \mathbf{u_2}(t-\tau) - \mathbf{i_2}(t-\tau) \\ \mathbf{g_1} \end{bmatrix} \quad (10)$$
$$\mathbf{i_1}(t) = -\mathbf{W} \cdot \mathbf{u_1}(t) + \mathbf{W} \cdot \mathbf{u_2}(t-\tau) - \mathbf{i_2}(t-\tau)$$

Set all the delays of transmission lines to be 1, then,

$$\begin{bmatrix} \mathbf{C_1} + \mathbf{W} & \mathbf{E_1} \\ \mathbf{F_1} & \mathbf{D_1} \end{bmatrix} \begin{bmatrix} \mathbf{u_1}^k \\ \mathbf{y_1}^k \end{bmatrix} = \begin{bmatrix} \mathbf{f_1} + \mathbf{W} \cdot \mathbf{u_2}^{k-1} - \mathbf{i_2}^{k-1} \\ \mathbf{g_1} \end{bmatrix} \quad (11)$$
$$\mathbf{i_1}^k = -\mathbf{W} \cdot \mathbf{u_1}^k + \mathbf{W} \cdot \mathbf{u_2}^{k-1} - \mathbf{i_2}^{k-1}$$

(11) is an SPD sparse linear system, and it could be solved by any sequential algorithms, such as cholesky factorization, algebraic multigrid method (AMG) and conjugate gradient method (CG).

Similarly, the analytic expression of sub-circuit 2 is obtained by merging (5) and (8):

$$\begin{bmatrix} \mathbf{C_2} & \mathbf{E_2} \\ \mathbf{F_2} & \mathbf{D_2} \end{bmatrix} \begin{bmatrix} \mathbf{u_2}(t) \\ \mathbf{y_2}(t) \end{bmatrix} = \begin{bmatrix} \mathbf{f_2} \\ \mathbf{g_2} \end{bmatrix} + \begin{bmatrix} \mathbf{i_2}(t) \\ 0 \end{bmatrix} \quad (12)$$
$$\mathbf{i_2}(t) + \mathbf{Z} \cdot \mathbf{u_2}(t) = -\mathbf{i_1}(t-\tau) + \mathbf{W} \cdot \mathbf{u_1}(t-\tau)$$

Eliminate $\mathbf{i_2}(t)$, and we obtain:

$$\begin{bmatrix} \mathbf{C_2} + \mathbf{W} & \mathbf{E_2} \\ \mathbf{F_2} & \mathbf{D_2} \end{bmatrix} \begin{bmatrix} \mathbf{u_2}(t) \\ \mathbf{y_2}(t) \end{bmatrix} = \begin{bmatrix} \mathbf{f_2} + \mathbf{W} \cdot \mathbf{u_1}(t-\tau) - \mathbf{i_1}(t-\tau) \\ \mathbf{g_2} \end{bmatrix} \quad (13)$$
$$\mathbf{i_2}(t) = -\mathbf{W} \cdot \mathbf{u_2}(t) + \mathbf{W} \cdot \mathbf{u_1}(t-\tau) - \mathbf{i_1}(t-\tau)$$

If all the delays of lines are 1, we obtain:

$$\begin{bmatrix} \mathbf{C_2} + \mathbf{W} & \mathbf{E_2} \\ \mathbf{F_2} & \mathbf{D_2} \end{bmatrix} \begin{bmatrix} \mathbf{u_2}^k \\ \mathbf{y_2}^k \end{bmatrix} = \begin{bmatrix} \mathbf{f_2} + \mathbf{W} \cdot \mathbf{u_1}^{k-1} - \mathbf{i_1}^{k-1} \\ \mathbf{g_2} \end{bmatrix} \quad (14)$$
$$\mathbf{i_2}^k = -\mathbf{W} \cdot \mathbf{u_2}^k + \mathbf{W} \cdot \mathbf{u_1}^{k-1} - \mathbf{i_1}^{k-1}$$

Later, we locate (11) on Processor 1, and locate (14) on Processor 2. After that, we set the initial values for boundary variables:

$$\mathbf{u_1^0} = \mathbf{u_2^0} = 0$$
$$\mathbf{i_1^0} = \mathbf{i_2^0} = 0$$

At last, we do the distributed computing.

## VII. CONVERGENCE THEORY

**Theorem 1** (Convergence Theory): Assume a linear resistor network is partitioned into $N$ sub-circuits by wire tearing, if the characteristic impedance matrix of the virtual transmission lines is symmetrical-positive-definite (SPD), then VTM converges at the answer to the resistor network.

This conclusion is supported by Lemma 2, and the mathematical proof is given in the appendix.

According to the proof of convergence theory, we know that the convergence speed of VTM is depending on the choice of the characteristic admittance matrix $\mathbf{W}$. So $\mathbf{W}$ is also called the preconditioner for VTM.

## VIII. PRECONDITIONING

In this section, we discuss the preconditioning techniques for VTM. We first define the input admittance matrix of a circuit or microwave network [18, 13]. From the view of numerical analysis, input admittance matrix is an alias of Schur complement matrix associated with the interface variables $\mathbf{u}$ [45, 68].

**Example 3**. Refer to (7), for Sub-circuit 1, the Schur complement matrix associated with the interface variables $\mathbf{u_1}$ is:

$$\mathbf{S}_1 = \mathbf{C}_1 - \mathbf{E}_1 \mathbf{D}_1^{-1} \mathbf{F}_1$$

$\mathbf{S}_1$ is also the input admittance matrix associated with the interface variables $\mathbf{u}_1$ in Sub-circuit 1.

Similarly, in Sub-circuit 2, the Schur complement matrix associated with the interface variables $\mathbf{u}_2$ is:

$$\mathbf{S}_2 = \mathbf{C}_2 - \mathbf{E}_2 \mathbf{D}_2^{-1} \mathbf{F}_2$$

$\mathbf{S}_2$ is also the input admittance matrix associated with the interface variables $\mathbf{u}_2$ in Sub-circuit 2.

Then, the voltage reflection matrix on the interface of Sub-circuit 1 is defined as:

$$\mathbf{T}_1 = (\mathbf{W} - \mathbf{S}_1)(\mathbf{W} + \mathbf{S}_1)^{-1}$$

The voltage reflection matrix on the interface of Sub-circuit 2 is:

$$\mathbf{T}_2 = (\mathbf{W} - \mathbf{S}_2)(\mathbf{W} + \mathbf{S}_2)^{-1}$$

According to the convergence proof in the appendix, the convergence factor of VTM is:

$$CF = \sqrt{\rho(\mathbf{T}_1 \mathbf{T}_2)}$$

Here $\rho(\mathbf{A})$ is the spectral radius of the square matrix $\mathbf{A}$.

Consequently, we know that the convergence speed of VTM is depending on the choice of the characteristic admittance matrix $\mathbf{W}$. This viewpoint is identical to the impedance matching technique in microwave engineering [13].

In the ideal case, if we set the preconditioner,

$$\mathbf{W} = \mathbf{S}_1 \text{ or } \mathbf{S}_2$$

Then,

$$\rho(\mathbf{T}_1 \mathbf{T}_2) = 0$$

As the result, only 1 iteration is needed to achieve convergence. Based on the knowledge of microwave network, we know that, if one port is precisely matched, there would be no energy reflection on this port and the network would settle down after 1 reflection.

In practice, it is impossible to obtain the accurate input admittance matrix of a large circuit, because the dimension of $\mathbf{D}_1$ or $\mathbf{D}_2$ might be very large, and it is impossible to obtain $\mathbf{D}_1^{-1}$ and $\mathbf{D}_2^{-1}$. So we need to figure out some practical way to approximate $\mathbf{W}$ to make $\mathbf{W} \approx \mathbf{S}_1$ or $\mathbf{S}_2$.

If the circuit is a resistor network, first we need to assure that $\mathbf{W}$ is SPD. Continue with Example 2, there are several ways to choose $\mathbf{W}$.

1. $\mathbf{W} = \alpha \cdot \mathbf{I}, \ \alpha > 0$. $\mathbf{I}$ is the identity matrix.

2. $\mathbf{W} = diag(\mathbf{C}_1) \ or \ diag(\mathbf{C}_2)$. We call it diagonal preconditioner.

3. $\mathbf{W} = \mathbf{C}_1 \ or \ \mathbf{C}_2$. This is called overlapped block preconditioner. It is similar to the overlapped block Jacobi method.

4. $\mathbf{W} = \alpha \cdot \mathbf{C}_1 \ or \ \alpha \cdot \mathbf{C}_2, \ \alpha > 0$. This is called weighted overlapped block preconditioner (WOB).

5. $\mathbf{W} \simeq \tilde{\mathbf{S}}_1 \ or \ \tilde{\mathbf{S}}_2$. $\tilde{\mathbf{S}}_1$ or $\tilde{\mathbf{S}}_2$ are called the partial input admittance matrix of the large circuit, which are defined as below:

$$\tilde{\mathbf{S}}_1 = \mathbf{C}_1 - \tilde{\mathbf{E}}_1 \tilde{\mathbf{D}}_1^{-1} \tilde{\mathbf{F}}_1$$

$$\tilde{\mathbf{S}}_2 = \mathbf{C}_2 - \tilde{\mathbf{E}}_2 \tilde{\mathbf{D}}_2^{-1} \tilde{\mathbf{F}}_2$$

Here $\begin{bmatrix} \mathbf{C}_1 & \tilde{\mathbf{E}}_1 \\ \tilde{\mathbf{F}}_1 & \tilde{\mathbf{D}}_1 \end{bmatrix}$ is a submatrix of $\begin{bmatrix} \mathbf{C}_1 & \mathbf{E}_1 \\ \mathbf{F}_1 & \mathbf{D}_1 \end{bmatrix}$. Similarly, $\begin{bmatrix} \mathbf{C}_2 & \tilde{\mathbf{E}}_2 \\ \tilde{\mathbf{F}}_2 & \tilde{\mathbf{D}}_2 \end{bmatrix}$ is a submatrix of $\begin{bmatrix} \mathbf{C}_2 & \mathbf{E}_2 \\ \mathbf{F}_2 & \mathbf{D}_2 \end{bmatrix}$.

This preconditioning technique is called Schur complement approximation method (SCA). Usually $\tilde{\mathbf{S}}_1$ or $\tilde{\mathbf{S}}_2$ is dense matrix, and we need to drop the small elements inside the matrix. After that, the fill-in pattern of $\mathbf{W}$ is similar to $\mathbf{C}_1$ or $\mathbf{C}_2$.

If the circuit is a general circuit with controlled sources, i.e. the linear system of circuit is unsymmetrical, the above 5 preconditioning techniques could also be used, but VTM might be unconvergent.

We have to say that, it is still an open problem to choose the preconditioner for the unsymmetrical linear system. In the next section, we illustrate a simple example to show the potential of VTM to solve the unsymmetrical system.

## IX. EXAMPLE

For the symmetrical linear system, a simple example could be found in [59]. Here we illustrate another example to solve the unsymmetrical linear system of the circuit. Fig. 5 proposes a tightly-coupled circuit with 4 operational amplifiers connected end-to-end. The linear system of this circuit is:

$$\begin{bmatrix} g_1 & g_{21} & 0 & 0 \\ 0 & g_2 & 0 & g_{42} \\ g_{13} & 0 & g_3 & 0 \\ 0 & 0 & g_{34} & g_4 \end{bmatrix} \begin{bmatrix} u_1 \\ u_2 \\ u_3 \\ u_4 \end{bmatrix} = \begin{bmatrix} I_1 \\ I_2 \\ I_3 \\ I_4 \end{bmatrix} \quad (15)$$

Here we set:

$$g_1 = g_2 = g_3 = g_4 = 1$$

$$g_{13} = g_{34} = g_{42} = -g_{21} = 10$$

The loop gain of this circuit is:

$$Gain = g_{13} g_{34} g_{42} g_{21} g_1^{-1} g_3^{-1} g_4^{-1} g_2^{-1} = -10^4$$

Then we partition this linear system by wire tearing,

$$\left[\begin{array}{cc|cc} g_1 & g_{21} & 0 & 0 \\ \hline 0 & g_2 & 0 & g_{42} \\ g_{13} & 0 & g_3 & 0 \\ \hline 0 & 0 & g_{34} & g_4 \end{array}\right] \begin{bmatrix} u_1 \\ u_2 \\ u_3 \\ u_4 \end{bmatrix} = \begin{bmatrix} I_1 \\ I_2 \\ I_3 \\ I_4 \end{bmatrix}$$

As the result, we obtain 2 sub-systems:

$$\begin{bmatrix} g_1 & g_{21} & 0 \\ 0 & g_{2a} & 0 \\ g_{13} & 0 & g_{3a} \end{bmatrix} \begin{bmatrix} u_1 \\ u_{2a} \\ u_{3a} \end{bmatrix} = \begin{bmatrix} I_1 \\ I_{2a} + i_{2a} \\ I_{3a} + i_{2b} \end{bmatrix} \quad (16)$$

$$\begin{bmatrix} g_{2b} & 0 & g_{42} \\ 0 & g_{3b} & 0 \\ 0 & g_{34} & g_4 \end{bmatrix} \begin{bmatrix} u_{2b} \\ u_{3b} \\ u_4 \end{bmatrix} = \begin{bmatrix} I_{2b} + i_{2b} \\ I_{3b} + i_{3b} \\ I_4 \end{bmatrix} \quad (17)$$

Here:

$$g_{2a} + g_{2b} = g_2, g_{3a} + g_{3b} = g_3$$
$$I_{2a} + I_{2b} = I_2, I_{3a} + I_{3b} = I_3$$

For simplicity, we set:

$$g_{2a} = g_{2b} = g_{3a} = g_{3b} = 0.5$$

The transmission delay equations of $T_2$ are:

$$\begin{aligned} i_{2a}(t) + W_2 \cdot u_{2a}(t) &= -i_{2b}(t-\tau) + W_2 \cdot u_{2b}(t-\tau) \\ i_{2b}(t) + W_2 \cdot u_{2b}(t) &= -i_{2a}(t-\tau) + W_2 \cdot u_{2a}(t-\tau) \end{aligned} \quad (18)$$

The transmission delay equations of $T_3$ are:

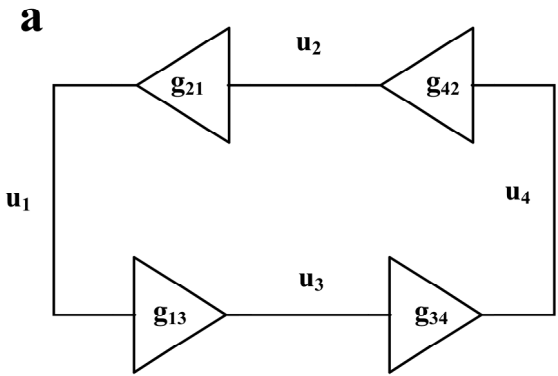
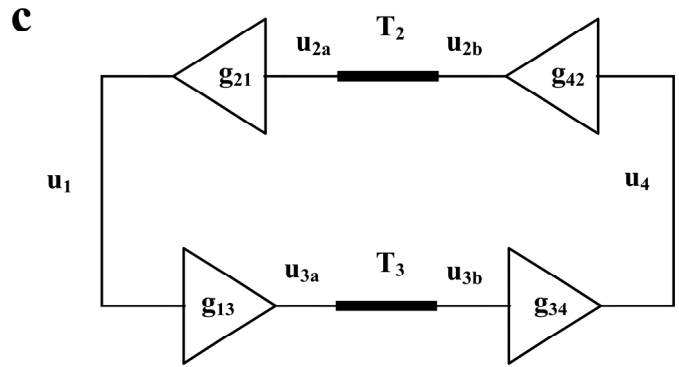
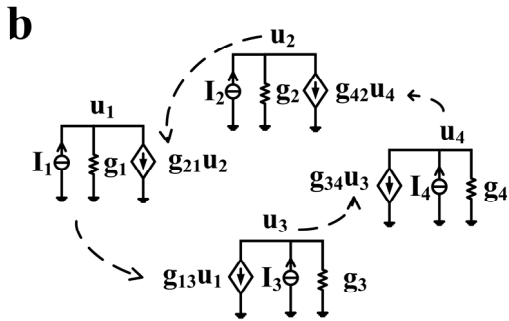
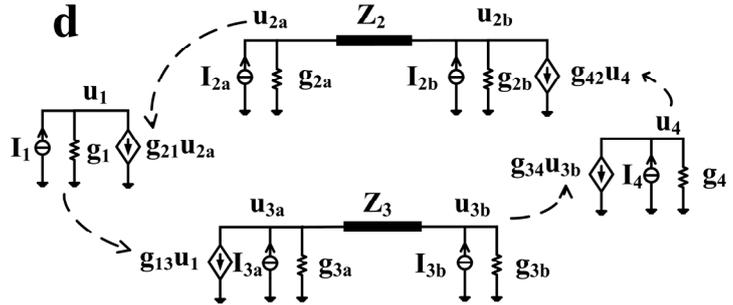
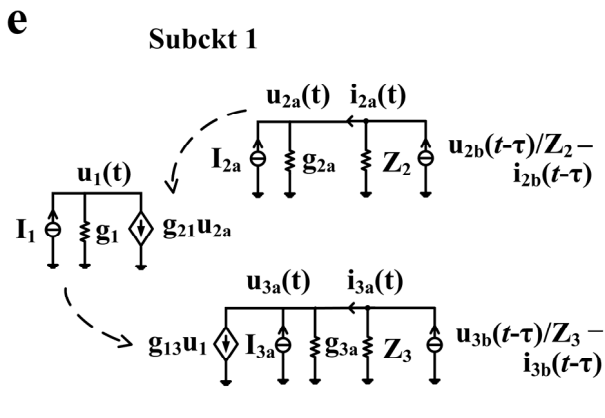
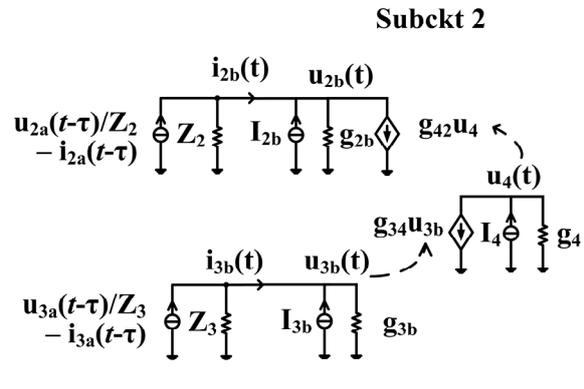

Fig. 5. Solving the unsymmetric linear system by VTM . (**a**) The original circuit with operational amplifiers. (**b**) Equivalent circuit of (**a**), with voltage controlled current sources (VCCS). (**c**) Partition the circuit by wire tearing. Node 2 and 3 is set to be the interfacial node. (**d**) Partition the equivalent circuit by wire tearing. $1/R_2 = 1/R_{2a} + 1/R_{2b}$. $I_2 = I_{2a} + I_{2b}$. $1/R_3 = 1/R_{3a} + 1/R_{3b}$. $I_3 = I_{3a} + I_{3b}$. (**e**) Replace the virtual transmission line with its equivalent circuit, and thus the original circuit is partitioned into two sub-circuits.

$$i_{3a}(t)+W_3 \cdot u_{3a}(t)=-i_{3b}(t-\tau)+W_3 \cdot u_{3b}(t-\tau)$$
$$i_{3b}(t)+W_3 \cdot u_{3b}(t)=-i_{3a}(t-\tau)+W_3 \cdot u_{3a}(t-\tau)$$
(19)

We set:
$$W_2 = 0.5, \; W_3 = 20000$$

Merge (16), (18) and (19), we obtain the linear system for Sub-circuit 1:
$$\begin{bmatrix} g_{2a} & 0 \\ g_{13}g_1^{-1}g_{21} & g_{3a} \end{bmatrix}\begin{bmatrix} u_{2a}(t) \\ u_{3a}(t) \end{bmatrix}=\begin{bmatrix} I_{2a}+i_{2a}(t) \\ I_{3a}-g_{13}g_1^{-1}I_1+i_{3a}(t) \end{bmatrix}$$
$$i_{2a}(t)+W_2 \cdot u_{2a}(t)=-i_{2b}(t-\tau)+W_2 \cdot u_{2b}(t-\tau)$$
$$i_{3a}(t)+W_3 \cdot u_{3a}(t)=-i_{3b}(t-\tau)+W_3 \cdot u_{3b}(t-\tau)$$
(20)

Similarly, we get the linear system for Sub-circuit 2:
$$\begin{bmatrix} g_{2b} & g_{42}g_4^{-1}g_{34} \\ 0 & g_{3b} \end{bmatrix}\begin{bmatrix} u_{2b}(t) \\ u_{3b}(t) \end{bmatrix}=\begin{bmatrix} I_{2b}-g_{42}g_4^{-1}I_4+i_{2b}(t) \\ I_{3b}+i_{3b}(t) \end{bmatrix}$$
$$i_{2b}(t)+W_2 \cdot u_{2b}(t)=-i_{2a}(t-\tau)+W_2 \cdot u_{2a}(t-\tau)$$
$$i_{3b}(t)+W_3 \cdot u_{3b}(t)=-i_{3a}(t-\tau)+W_3 \cdot u_{3a}(t-\tau)$$
(21)

Finally we solve this example on 2 processors by VTM, and the computing result is compared with a number of well-known stationary iterative algorithms, e.g. Jacobi, Gauss Seidel (GS), Successive Overrelaxation Method (SOR), Symmetric SOR (SSOR), Block Jacobi (BJ) [45]. Here the global iterative index $k$ increases to $k+1$ until all the variables are updated.

As shown in Fig. 6, the traditional stationary algorithms are unconvergent for this circuit, but VTM is convergent. The key is to choose proper characteristic admittances for the virtual transmission lines.

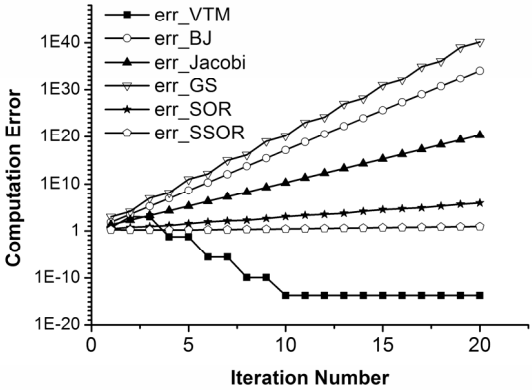

Fig. 6. Comparison of the convergence speed of iterative algorithms for resistor network.

According to this example, we conclude that there exists a parallel between the stability of VTM and the stability of distributed linear circuit (or microwave network). The traditional techniques to stabilize a distributed circuit might be transplanted to stabilize VTM.

TABLE I
SPARSE MATRICES PROPERTY

| Name | Interface nodes | Total nodes | Property | Source |
|---|---|---|---|---|
| Grid2d | 100 | 10000 | SPD | MESHPART |
| Grid3d | 900 | 27000 | SPD | H. Qian |
| Fv2 | 129 | 9801 | SPD | UFpack |
| Wang | 870 | 26064 | Unsymm. | UFpack |
| Pchip | 13 | 2298 | Unsymm. | MCNC'90 |

TABLE II
ITERATION NUMBER TO ACHIEVE CONVERGENCE

| Method | Grid2d | Grid3d | Fv2 | Wang | Pchip |
|---|---|---|---|---|---|
| Jacobi | 27759 | 5238 | 212 | 50773 | -- |
| GS | 18373 | 2950 | 109 | 27423 | -- |
| SGS | 11773 | 1571 | 79 | 26842 | -- |
| SOR | 14253 | 2049 | 66 | 16289 | -- |
| SSOR | 8839 | 1073 | 74 | 29135 | -- |
| BJ | 966 | 227 | 50 | 563 | 952 |
| OBJ | 490 | 114 | 23 | 287 | 162 |
| VTM_Diag | 963 | 335 | 42 | 854 | 160 |
| VTM_OB | 491 | 115 | 23 | 287 | 161 |
| VTM_WOB | 296 | 68 | 22 | 175 | -- |
| VTM_SCE | 275 | 75 | 14 | 186 | -- |

However, for the large unsymmetrical linear system, we have not yet figure out an effective method to choose **W**. The convergence property of VTM is depending on the property of linear circuit, as well as the preconditioner **W**.

X. NUMERICAL EXPERIMENTS

In this section we compare VTM with the traditional stationary algorithms for both the symmetrical matrix and unsymmetrical matrix. We use MATLAB and SIMULINK as our test environment [53, 54], and use METIS and MESHPART as our partitioning methods [28, 51, 19].

The test matrices are described in Table 1. Grid2d and Grid3d are structured resistor networks [51, 43]. Fv2 is a 2D mesh constructed by finite element method (FEM) [14]. Pchip is extracted from a MOS integrated circuit of MCNC'90 testbench [52]. Wang is a sparse matrix from semiconductor device simulation [14].

Table 2 illustrates the computing results of the stationery iterative algorithms. BJ, OBJ and VTM are tested on 2 processors. Jacobi, GS, Symmetrical GS (SGS), SOR and SSOR are tested on 1 processor.

VTM_Diag means VTM with diagonal preconditioner. VTM_OB represents VTM with overlapped block preconditioner. VTM_WOB is VTM with weighted overlapped block preconditioner. VTM_SCA means VTM with Schur complement approximation preconditioner. All these preconditioning methods were previously described in Section 8.

According to Table 2, we conclude that VTM is convergent to solve the SPD system, and its convergence speed is fast when using the SCA or WOB preconditioner. The convergence speed of OBJ and VTM_OB are same. For the unsymmetrical linear systems, VTM might be unconvergent.

At last, we test VTM on a various number of processors, as shown in Fig. 7. In this case, we use Grid3d as the testbench. The test result indicates that the convergence speed of VTM is slightly relative to the number of processors.

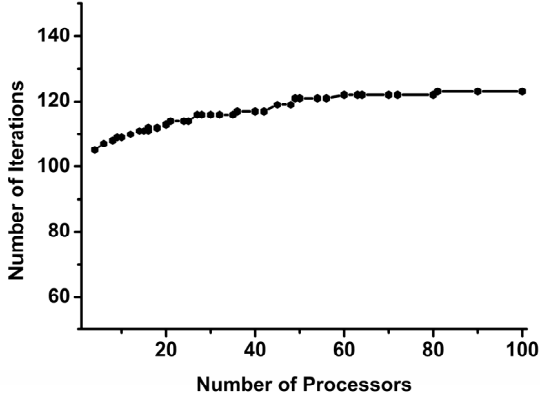

Fig. 7. Test VTM on a number of processors.

## XI. CONCLUSION AND FUTURE WORK

In this paper, we introduce VTM to solve the sparse linear system of circuit in parallel. VTM is a distributed and stationary iterative numerical algorithm, and it is efficient to solve the resistor networks.

VTM proves many insightful conclusions linking numerics and electrics. The physical background of VTM is electric circuit, and this makes this algorithm different from the traditional numerical algorithms. As presented in this paper, we have borrowed many concepts from circuit theory to describe this algorithm.

According to the simple example presented in Section 9, we conclude that there exists similarity between the stability of VTM and the stability of distributed linear circuit (or microwave network).

The convergence speed of VTM is highly depending on the characteristic admittance matrix of the virtual transmission lines, i.e. its preconditioner $\mathbf{W}$. A number of preconditioning techniques are implemented to accelerate VTM. Experiments indicate that, if $\mathbf{W}$ is properly chosen, the performance of VTM would be appreciable; if not, VTM would be plain. Nevertheless, the computation cost to design the preconditioner $\mathbf{W}$ should be aware of [5].

For the SPD linear system, we have proved that VTM is convergent; for the large unsymmetrical linear system, we have not yet figure out an effective way to choose the characteristic admittance matrix to guarantee convergence. This is an open problem.

VTM could also be used to solve the nonlinear system, as we pointed out in [59]. However, the convergence analysis of nonlinear system is more complicated than the linear system. This problem should be further studied.

If the delays of virtual transmission lines are different and stochastic, VTM would turn to be a fully asynchronous and chaotic numerical algorithm [58]. If we replace each variable by a piece of waveform, VTM would become a waveform-relaxation algorithm [60].

To simulate the post-layout integrated circuit in parallel, we have figured out another method to distributedly solve the nonlinear delay differential equations extracted from the distributed circuits [61].

## APPENDIX

Here we prove that, if the graph of the resistor network is 2-partete, VTM converges.

**Lemma 3**: Assume a linear resistor network is partitioned into 2 sub-circuits by level-one wire tearing, if the characteristic impedance matrix $\mathbf{Z}$ of the virtual transmission lines is SPD, VTM converges at the answer to the resistor network.

Proof:

First, we consider the resistor network with inner nodes, and eliminate these inner nodes by Schur complement technique. Then, we prove Lemma 3 for the resistor network without inner nodes.

The sparse linear system of the resistor network is:

$$\begin{bmatrix} \mathbf{C} & \mathbf{E}_1 & \mathbf{E}_2 \\ \mathbf{F}_1 & \mathbf{D}_1 & \mathbf{0} \\ \mathbf{F}_2 & \mathbf{0} & \mathbf{D}_2 \end{bmatrix} \begin{bmatrix} \mathbf{u} \\ \mathbf{y}_1 \\ \mathbf{y}_2 \end{bmatrix} = \begin{bmatrix} \mathbf{f} \\ \mathbf{g}_1 \\ \mathbf{g}_2 \end{bmatrix}$$

Eliminate the inner variables $\mathbf{y}_1$ and $\mathbf{y}_2$, we obtain:

$$\left( \mathbf{C} - \mathbf{E}_1 \mathbf{D}_1^{-1} \mathbf{F}_1 - \mathbf{E}_2 \mathbf{D}_2^{-1} \mathbf{F}_2 \right) \mathbf{u}$$
$$= \mathbf{f} - \mathbf{E}_1 \mathbf{D}_1^{-1} \mathbf{g}_1 - \mathbf{E}_2 \mathbf{D}_2^{-1} \mathbf{g}_2$$

$$\mathbf{y}_1 = -\mathbf{D}_1^{-1} \mathbf{F}_1 \mathbf{u} + \mathbf{D}_1^{-1} \mathbf{g}_1$$

$$\mathbf{y}_2 = -\mathbf{D}_2^{-1} \mathbf{F}_2 \mathbf{u} + \mathbf{D}_2^{-1} \mathbf{g}_2$$

Set,

$$\mathbf{S} = \mathbf{C} - \mathbf{E}_1 \mathbf{D}_1^{-1} \mathbf{F}_1 - \mathbf{E}_2 \mathbf{D}_2^{-1} \mathbf{F}_2,$$
$$\mathbf{r} = \mathbf{f} - \mathbf{E}_1 \mathbf{D}_1^{-1} \mathbf{g}_1 - \mathbf{E}_2 \mathbf{D}_2^{-1} \mathbf{g}_2,$$

then,

$$\mathbf{S} \cdot \mathbf{u} = \mathbf{r} \quad (22)$$

$\mathbf{S}$ is called the Schur complement matrix associated with the interface variables $\mathbf{u}$.

**Lemma 4**: If $\mathbf{A} = \begin{bmatrix} \mathbf{C} & \mathbf{E} \\ \mathbf{F} & \mathbf{D} \end{bmatrix}$ is SPD, then the Schur complement matrix $\mathbf{S} = \mathbf{C} - \mathbf{E}\mathbf{D}^{-1}\mathbf{F}$ is SPD.

If the circuit is split into two sub-circuits by level-one wire tearing, the system of Sub-circuit 1 is (7):

$$\begin{bmatrix} \mathbf{C}_1 & \mathbf{E}_1 \\ \mathbf{F}_1 & \mathbf{D}_1 \end{bmatrix} \begin{bmatrix} \mathbf{u}_1 \\ \mathbf{y}_1 \end{bmatrix} = \begin{bmatrix} \mathbf{f}_1 \\ \mathbf{g}_1 \end{bmatrix} + \begin{bmatrix} \mathbf{i}_1 \\ \mathbf{0} \end{bmatrix}$$

Eliminate the inner variables $\mathbf{y}_1$:

$$(C_1 - E_1 D_1^{-1} F_1) u_1 = f_1 - E_1 D_1^{-1} g_1 + i_1$$

$$y_1 = -D_1^{-1} F_1 u_1 + D_1^{-1} g_1$$

Set $S_1 = C_1 - E_1 D_1^{-1} F_1$, $r_1 = f_1 - E_1 D_1^{-1} g_1$, then,

$$S_1 \cdot u_1 = r_1 + i_1 \tag{23}$$

$S_1$ is called the Schur complement matrix associated with the interface variables $u_1$ in Sub-circuit 1.

The system of Sub-circuit 2 is (8):

$$\begin{bmatrix} C_2 & E_2 \\ F_2 & D_2 \end{bmatrix} \begin{bmatrix} u_2 \\ y_2 \end{bmatrix} = \begin{bmatrix} f_2 \\ g_2 \end{bmatrix} + \begin{bmatrix} i_2 \\ 0 \end{bmatrix}$$

Eliminate the inner variables $y_2$:

$$(C_2 - E_2 D_2^{-1} F_2) u_2 = f_2 - E_2 D_2^{-1} g_2 + i_2$$

$$y_2 = -D_2^{-1} F_2 u_2 + D_2^{-1} g_2$$

Set $S_2 = C_2 - E_2 D_2^{-1} F_2$, $r_2 = f_2 - E_2 D_2^{-1} g_2$, then,

$$S_2 \cdot u_2 = r_2 + i_2 \tag{24}$$

$S_2$ is called the Schur complement matrix associated with the interface variables $u_2$ in Sub-circuit 2.

Consequently, we have:

$$\begin{aligned} S_1 + S_2 &= C_1 - E_1 D_1^{-1} F_1 + C_2 - E_2 D_2^{-1} F_2 \\ &= C - E_1 D_1^{-1} F_1 - E_2 D_2^{-1} F_2 \\ &= S \end{aligned}$$

$$\begin{aligned} r_1 + r_2 &= f_1 - E_1 D_1^{-1} g_1 + f_2 - E_2 D_2^{-1} g_2 \\ &= f - E_1 D_1^{-1} g_1 - E_2 D_2^{-1} g_2 \\ &= r \end{aligned}$$

Until now, we have eliminated all the inner nodes in the resistor network by Schur complement technique, and all the nodes are interfacial nodes. The linear system of (22) is partitioned into two sub-systems of (23) and (24) by wire tearing.

The iterative formula is Transmission Delay Equations (5),

$$i_1^k + W \cdot u_1^k = -i_2^{k-1} + W \cdot u_2^{k-1}$$

$$i_2^k + W \cdot u_2^k = -i_1^{k-1} + W \cdot u_1^{k-1}$$

Here $W$ is the characteristic admittance matrix of the transmission lines. $W$ is SPD. $Z = W^{-1}$.

Eliminate $i_1^k$ and $i_2^k$ by combine (23), (24) and (5), we find,

$$W \cdot u_1^k + (S_1 u_1^k - r_1) = W \cdot u_2^{k-1} - (S_2 u_2^{k-1} - r_2)$$

$$(W + S_1) u_1^k - r_1 = (W - S_2) u_2^{k-1} + r_2$$

Thus,

$$u_1^k = (W + S_1)^{-1} (W - S_2) u_2^{k-1} \tag{25}$$
$$+ (W + S_1)^{-1} r$$

Similarly, we obtain:

$$u_2^k = (W + S_2)^{-1} (W - S_1) u_1^{k-1} \tag{26}$$
$$+ (W + S_2)^{-1} r$$

Merge (25) and (26), we come to know the iterative formula for $u_1^k$,

$$\begin{aligned} u_1^k &= (W + S_1)^{-1} (W - S_2) \\ &\quad \times (W + S_2)^{-1} (W - S_1) u_1^{k-2} \\ &\quad + (W + S_1)^{-1} (W - S_2)(W + S_2)^{-1} r \\ &\quad + (W + S_1)^{-1} r \end{aligned} \tag{27}$$

Set:

$$T_1 = (W - S_1)(W + S_1)^{-1}$$

$$T_2 = (W - S_2)(W + S_2)^{-1}$$

Then,

$$\begin{aligned} u_1^k &= (W - S_1)^{-1} T_1 T_2 (W - S_1) u_1^{k-2} \\ &\quad + (W - S_1)^{-1} T_1 T_2 \cdot r \\ &\quad + (W - S_1)^{-1} T_1 \cdot r \end{aligned} \tag{28}$$

Reformat (28) as (29):

$$(W - S_1) u_1^k = T_1 T_2 (W - S_1) u_1^{k-2} \tag{29}$$
$$+ T_1 T_2 \cdot r + T_1 \cdot r$$

If we make use a new variable:

$$a_1^k = (W - S_1) u_1^k$$

then, (29) could be expressed as (30):

$$a_1^k = T_1 T_2 \cdot a_1^{k-2} + h_1 \tag{30}$$

where $h_1 = T_1 \cdot T_2 \cdot r + T_1 \cdot r$.

In the following text, we will prove that (30) is convergent. Here we use $Z^{-1}$ instead of $W$.

$$\begin{aligned} T_1 &= (Z^{-1} - S_1)(Z^{-1} + S_1)^{-1} \\ &= Z^{-1} (I - Z \cdot S_1)(I + Z \cdot S_1)^{-1} Z \end{aligned} \tag{31}$$

$$\begin{aligned} T_2 &= (Z^{-1} - S_2)(Z^{-1} + S_2)^{-1} \\ &= Z^{-1} (I - Z \cdot S_2)(I + Z \cdot S_2)^{-1} Z \end{aligned} \tag{32}$$

**Lemma 5**: If $Z$ is an SPD matrix of dimension $n$, then there exists a matrix $\sqrt{Z}$, which satisfies: $\sqrt{Z}^T \sqrt{Z} = Z$.

Proof: Because $Z$ is SPD,

$$Z = Q^T \Lambda Q = Q^T \sqrt{\Lambda} \sqrt{\Lambda} Q = \sqrt{Z}^T \sqrt{Z}$$

where $\mathbf{Q}^T\mathbf{Q} = \mathbf{I}$. $\mathbf{I}$ is the identity matrix.
$$\mathbf{\Lambda} = diag(\alpha_1, \alpha_2, \cdots, \alpha_n), \alpha_i \in \mathbb{R}^+.$$
$$\sqrt{\mathbf{\Lambda}} = diag(\sqrt{\alpha_1}, \sqrt{\alpha_2}, \cdots, \sqrt{\alpha_n}).$$
$$\sqrt{\mathbf{Z}} = \sqrt{\mathbf{\Lambda}}\mathbf{Q}.$$
End of proof [25].

**Lemma 6**: If both $\mathbf{Z}$ and $\mathbf{S}$ are SPD matrix of dimension $n$, then $\mathbf{Z}\cdot\mathbf{S}$ has the same eigenvalues as $\sqrt{\mathbf{Z}}\cdot\mathbf{S}\cdot\sqrt{\mathbf{Z}}^T$.

Proof: Because $\mathbf{S}$ and $\mathbf{Z}$ are SPD, $\mathbf{Z}\cdot\mathbf{S}$ is positive-definite, and $\sqrt{\mathbf{Z}}^T\mathbf{S}\sqrt{\mathbf{Z}}$ is SPD.

Assume $\sqrt{\mathbf{Z}}\cdot\mathbf{S}\cdot\sqrt{\mathbf{Z}}^T = \mathbf{Q}\mathbf{T}\mathbf{Q}^T$, where $\mathbf{Q}\mathbf{Q}^T = \mathbf{I}$, $\mathbf{T} = diag(t_1, t_2, \cdots, t_n)$, $t_i > 0, i = 1, 2, \cdots, n$. Then,

$$\mathbf{Z}\cdot\mathbf{S} = \sqrt{\mathbf{Z}}^T \cdot \sqrt{\mathbf{Z}} \cdot \mathbf{S} \cdot \sqrt{\mathbf{Z}}^T \cdot \left(\sqrt{\mathbf{Z}}^T\right)^{-1}$$
$$= \sqrt{\mathbf{Z}}^T \cdot \mathbf{Q}\mathbf{T}\mathbf{Q}^T \cdot \left(\sqrt{\mathbf{Z}}^T\right)^{-1}$$
$$= \left(\sqrt{\mathbf{Z}}^T\mathbf{Q}\right)\cdot\mathbf{T}\cdot\left(\sqrt{\mathbf{Z}}^T\mathbf{Q}\right)^{-1}$$

End of proof.

According to Lemma 5 and 6, we come to know that,
$$\sqrt{\mathbf{Z}}\cdot\mathbf{S}_1\cdot\sqrt{\mathbf{Z}}^T = \mathbf{Q}_1\mathbf{\Lambda}_1\mathbf{Q}_1^T$$
where $\mathbf{\Lambda}_1 = diag(\lambda_1, \lambda_2, \cdots, \lambda_n)$, $\lambda_i > 0$, $i = 1, \cdots, n$,
$\mathbf{Q}_1\mathbf{Q}_1^T = \mathbf{I}$. $\mathbf{Z}\cdot\mathbf{S}_1 = \left(\sqrt{\mathbf{Z}}^T\mathbf{Q}_1\right)\mathbf{\Lambda}_1\left(\sqrt{\mathbf{Z}}^T\mathbf{Q}_1\right)^{-1}$.

Similarly, we obtain,
$$\sqrt{\mathbf{Z}}\cdot\mathbf{S}_2\cdot\sqrt{\mathbf{Z}}^T = \mathbf{Q}_2\mathbf{\Lambda}_2\mathbf{Q}_2^T$$
where $\mathbf{\Lambda}_2 = diag(\mu_1, \mu_2, \cdots, \mu_n)$, $\mu_i > 0$, $i = 1, \cdots, n$,
$\mathbf{Q}_2\mathbf{Q}_2^T = \mathbf{I}$. $\mathbf{Z}\cdot\mathbf{S}_2 = \left(\sqrt{\mathbf{Z}}^T\mathbf{Q}_2\right)\mathbf{\Lambda}_2\left(\sqrt{\mathbf{Z}}^T\mathbf{Q}_2\right)^{-1}$.

Then, (31) could be decomposed as below:

$$\mathbf{T}_1 = \mathbf{Z}^{-1}(\mathbf{I} - \mathbf{Z}\cdot\mathbf{S}_1)(\mathbf{I} + \mathbf{Z}\cdot\mathbf{S}_1)^{-1}\mathbf{Z}$$
$$= \mathbf{Z}^{-1}\left(\begin{array}{c}\left(\sqrt{\mathbf{Z}}^T\mathbf{Q}_1\right)\left(\sqrt{\mathbf{Z}}^T\mathbf{Q}_1\right)^{-1}\\ -\left(\sqrt{\mathbf{Z}}^T\mathbf{Q}_1\right)\mathbf{\Lambda}_1\left(\sqrt{\mathbf{Z}}^T\mathbf{Q}_1\right)^{-1}\end{array}\right)$$
$$\times\left(\begin{array}{c}\left(\sqrt{\mathbf{Z}}^T\mathbf{Q}_1\right)\left(\sqrt{\mathbf{Z}}^T\mathbf{Q}_1\right)^{-1}\\ +\left(\sqrt{\mathbf{Z}}^T\mathbf{Q}_1\right)\mathbf{\Lambda}_1\left(\sqrt{\mathbf{Z}}^T\mathbf{Q}_1\right)^{-1}\end{array}\right)^{-1}\mathbf{Z}$$
$$= \mathbf{Z}^{-1}\left(\sqrt{\mathbf{Z}}^T\mathbf{Q}_1\right)(\mathbf{I}-\mathbf{\Lambda}_1)\left(\sqrt{\mathbf{Z}}^T\mathbf{Q}_1\right)^{-1}$$
$$\times\left(\sqrt{\mathbf{Z}}^T\mathbf{Q}_1\right)(\mathbf{I}+\mathbf{\Lambda}_1)^{-1}\left(\sqrt{\mathbf{Z}}^T\mathbf{Q}_1\right)^{-1}\mathbf{Z}$$
$$= \mathbf{Z}^{-1}\left(\sqrt{\mathbf{Z}}^T\mathbf{Q}_1\right)(\mathbf{I}-\mathbf{\Lambda}_1)(\mathbf{I}+\mathbf{\Lambda}_1)^{-1}\left(\sqrt{\mathbf{Z}}^T\mathbf{Q}_1\right)^{-1}\mathbf{Z}$$
$$= \mathbf{Z}^{-1}\left(\sqrt{\mathbf{Z}}^T\mathbf{Q}_1\right)\cdot diag\left(\frac{1-\lambda_1}{1+\lambda_1}, \cdots, \frac{1-\lambda_n}{1+\lambda_n}\right)\cdot\left(\sqrt{\mathbf{Z}}^T\mathbf{Q}_1\right)^{-1}\mathbf{Z}$$

Similarly, (32) could be decomposed as:
$$\mathbf{T}_2 = \mathbf{Z}^{-1}(\mathbf{I}-\mathbf{Z}\cdot\mathbf{S}_2)(\mathbf{I}+\mathbf{Z}\cdot\mathbf{S}_2)^{-1}\mathbf{Z}$$
$$= \mathbf{Z}^{-1}\left(\sqrt{\mathbf{Z}}^T\mathbf{Q}_2\right)(\mathbf{I}-\mathbf{\Lambda}_2)(\mathbf{I}+\mathbf{\Lambda}_2)^{-1}\left(\sqrt{\mathbf{Z}}^T\mathbf{Q}_2\right)^{-1}\mathbf{Z}$$
$$= \mathbf{Z}^{-1}\left(\sqrt{\mathbf{Z}}^T\mathbf{Q}_2\right)\cdot diag\left(\frac{1-\mu_1}{1+\mu_1}, \cdots, \frac{1-\mu_n}{1+\mu_n}\right)\cdot\left(\sqrt{\mathbf{Z}}^T\mathbf{Q}_2\right)^{-1}\mathbf{Z}$$

The spectral radius of $\mathbf{T}_1\mathbf{T}_2$ is defined as:
$$\rho(\mathbf{T}_1\mathbf{T}_2) = \lim_{k\to\infty}\left\|(\mathbf{T}_1\mathbf{T}_2)^k\right\|^{1/k}$$

So we first calculate $\left\|(\mathbf{T}_1\mathbf{T}_2)^k\right\|$. Here $\|\mathbf{A}\|$ is the spectral norm of the square matrix $\mathbf{A}$.

$$\left\|(\mathbf{T_1T_2})^k\right\|$$

$$=\left\|\begin{bmatrix}\mathbf{Z}^{-1}\left(\sqrt{\mathbf{Z}}^\mathbf{T}\mathbf{Q}_1\right)(\mathbf{I}-\mathbf{\Lambda}_1)(\mathbf{I}+\mathbf{\Lambda}_1)^{-1}\left(\sqrt{\mathbf{Z}}^\mathbf{T}\mathbf{Q}_1\right)^{-1}\mathbf{Z}\\\times\mathbf{Z}^{-1}\left(\sqrt{\mathbf{Z}}^\mathbf{T}\mathbf{Q}_2\right)(\mathbf{I}-\mathbf{\Lambda}_2)(\mathbf{I}+\mathbf{\Lambda}_2)^{-1}\left(\sqrt{\mathbf{Z}}^\mathbf{T}\mathbf{Q}_2\right)^{-1}\mathbf{Z}\end{bmatrix}^k\right\|$$

$$=\left\|\sqrt{\mathbf{Z}}^{-1}\begin{bmatrix}\mathbf{Q}_1(\mathbf{I}-\mathbf{\Lambda}_1)(\mathbf{I}+\mathbf{\Lambda}_1)^{-1}\mathbf{Q}_1^\mathbf{T}\\\times\mathbf{Q}_2(\mathbf{I}-\mathbf{\Lambda}_2)(\mathbf{I}+\mathbf{\Lambda}_2)^{-1}\mathbf{Q}_2^\mathbf{T}\end{bmatrix}^k\sqrt{\mathbf{Z}}\right\|$$

$$\le\left\|\sqrt{\mathbf{Z}}^{-1}\right\|\cdot\left\|\begin{matrix}\mathbf{Q}_1(\mathbf{I}-\mathbf{\Lambda}_1)(\mathbf{I}+\mathbf{\Lambda}_1)^{-1}\mathbf{Q}_1^\mathbf{T}\\\times\mathbf{Q}_2(\mathbf{I}-\mathbf{\Lambda}_2)(\mathbf{I}+\mathbf{\Lambda}_2)^{-1}\mathbf{Q}_2^\mathbf{T}\end{matrix}\right\|^k\cdot\left\|\sqrt{\mathbf{Z}}\right\|$$

$$\le\left\|\sqrt{\mathbf{Z}}^{-1}\right\|\cdot\begin{bmatrix}\|\mathbf{Q}_1\|\cdot\|(\mathbf{I}-\mathbf{\Lambda}_1)(\mathbf{I}+\mathbf{\Lambda}_1)^{-1}\|\cdot\|\mathbf{Q}_1^\mathbf{T}\|\\\times\|\mathbf{Q}_2\|\cdot\|(\mathbf{I}-\mathbf{\Lambda}_2)(\mathbf{I}+\mathbf{\Lambda}_2)^{-1}\|\cdot\|\mathbf{Q}_2^\mathbf{T}\|\end{bmatrix}^k$$
$$\times\left\|\sqrt{\mathbf{Z}}\right\|$$

$$\le\left\|\sqrt{\mathbf{Z}}^{-1}\right\|\cdot\begin{bmatrix}\|(\mathbf{I}-\mathbf{\Lambda}_1)(\mathbf{I}+\mathbf{\Lambda}_1)^{-1}\|\cdot\\\times\|(\mathbf{I}-\mathbf{\Lambda}_2)(\mathbf{I}+\mathbf{\Lambda}_2)^{-1}\|\end{bmatrix}^k\cdot\left\|\sqrt{\mathbf{Z}}\right\|$$

$$=\left\|\sqrt{\mathbf{Z}}^{-1}\right\|\cdot\left\|diag\left(\frac{1-\lambda_1}{1+\lambda_1},\cdots,\frac{1-\lambda_n}{1+\lambda_n}\right)\right\|^k$$
$$\times\left\|diag\left(\frac{1-\mu_1}{1+\mu_1},\cdots,\frac{1-\mu_n}{1+\mu_n}\right)\right\|^k\cdot\left\|\sqrt{\mathbf{Z}}\right\|$$

$$\le\left\|\sqrt{\mathbf{Z}}^{-1}\right\|\cdot\max{}^k\left(\left\|\frac{1-\lambda_1}{1+\lambda_1}\right\|,\cdots,\left\|\frac{1-\lambda_n}{1+\lambda_n}\right\|\right)$$
$$\times\max{}^k\left(\left\|\frac{1-\mu_1}{1+\mu_1}\right\|,\cdots,\left\|\frac{1-\mu_n}{1+\mu_n}\right\|\right)\cdot\left\|\sqrt{\mathbf{Z}}\right\|$$

As the result,

$$\rho(\mathbf{T_1T_2})=\lim_{k\to\infty}\left\|(\mathbf{T_1T_2})^k\right\|^{1/k}$$
$$\le\max\left(\left\|\frac{1-\lambda_1}{1+\lambda_1}\right\|,\cdots,\left\|\frac{1-\lambda_n}{1+\lambda_n}\right\|\right)$$
$$\times\max\left(\left\|\frac{1-\mu_1}{1+\mu_1}\right\|,\cdots,\left\|\frac{1-\mu_n}{1+\mu_n}\right\|\right)$$
$$=\rho(\mathbf{T_1})\rho(\mathbf{T_2})$$

Because $\lambda_i > 0$, $i = 1,\cdots,n$,

$$\rho(\mathbf{T_1})=\max\left(\left\|\frac{1-\lambda_1}{1+\lambda_1}\right\|,\cdots,\left\|\frac{1-\lambda_n}{1+\lambda_n}\right\|\right)<1;$$

Because $\mu_i > 0$, $i = 1,\cdots,n$,

$$\rho(\mathbf{T_2})=\max\left(\left\|\frac{1-\mu_1}{1+\mu_1}\right\|,\cdots,\left\|\frac{1-\mu_n}{1+\mu_n}\right\|\right)<1.$$

Consequently,

$$\rho(\mathbf{T_1T_2})<1$$

So we conclude that VTM converges for 2-partite resistor network [45].

According to (27), when this algorithm is convergent, $\mathbf{u}_1^k = \mathbf{u}_1^{k-2} = \mathbf{u}_1^\infty$, then,

$$\mathbf{u}_1^k=(\mathbf{W}+\mathbf{S_1})^{-1}(\mathbf{W}-\mathbf{S_2})$$
$$\times(\mathbf{W}+\mathbf{S_2})^{-1}(\mathbf{W}-\mathbf{S_1})\mathbf{u}_1^{k-2}$$
$$+(\mathbf{W}+\mathbf{S_1})^{-1}(\mathbf{W}-\mathbf{S_2})(\mathbf{W}+\mathbf{S_2})^{-1}\mathbf{r}$$
$$+(\mathbf{W}+\mathbf{S_1})^{-1}\mathbf{r}$$

$$\mathbf{u}_1^\infty=\left[\mathbf{I}-(\mathbf{W}+\mathbf{S_1})^{-1}(\mathbf{W}-\mathbf{S_2})(\mathbf{W}+\mathbf{S_2})^{-1}(\mathbf{W}-\mathbf{S_1})\right]^{-1}$$
$$\times\left[(\mathbf{W}+\mathbf{S_1})^{-1}+(\mathbf{W}+\mathbf{S_1})^{-1}(\mathbf{W}-\mathbf{S_2})(\mathbf{W}+\mathbf{S_2})^{-1}\right]\cdot\mathbf{r}$$

$$=\left[(\mathbf{W}+\mathbf{S_1})-(\mathbf{W}-\mathbf{S_2})(\mathbf{W}+\mathbf{S_2})^{-1}(\mathbf{W}-\mathbf{S_1})\right]^{-1}$$
$$\times\left[\mathbf{I}+(\mathbf{W}-\mathbf{S_2})(\mathbf{W}+\mathbf{S_2})^{-1}\right]\cdot\mathbf{r}$$

$$=\begin{bmatrix}(\mathbf{I}-(\mathbf{W}-\mathbf{S_2})(\mathbf{W}+\mathbf{S_2})^{-1})\mathbf{W}+\\(\mathbf{I}+(\mathbf{W}-\mathbf{S_2})(\mathbf{W}+\mathbf{S_2})^{-1})\mathbf{S_1}\end{bmatrix}^{-1}$$
$$\times\left[\mathbf{I}+(\mathbf{W}-\mathbf{S_2})(\mathbf{W}+\mathbf{S_2})^{-1}\right]\cdot\mathbf{r}$$

$$=\begin{bmatrix}(\mathbf{I}+(\mathbf{W}-\mathbf{S_2})(\mathbf{W}+\mathbf{S_2})^{-1})^{-1}\\\times(\mathbf{I}-(\mathbf{W}-\mathbf{S_2})(\mathbf{W}+\mathbf{S_2})^{-1})\mathbf{W}+\mathbf{S_1}\end{bmatrix}^{-1}\mathbf{r}$$

$$=\begin{bmatrix}(\mathbf{W}+\mathbf{S_2})((\mathbf{W}+\mathbf{S_2})+(\mathbf{W}-\mathbf{S_2}))^{-1}\\\times((\mathbf{W}+\mathbf{S_2})-(\mathbf{W}-\mathbf{S_2}))(\mathbf{W}+\mathbf{S_2})^{-1}\mathbf{W}+\mathbf{S_1}\end{bmatrix}^{-1}\mathbf{r}$$

$$=\left[(\mathbf{W}+\mathbf{S_2})(2\mathbf{W})^{-1}(2\mathbf{S_2})(\mathbf{W}+\mathbf{S_2})^{-1}\mathbf{W}+\mathbf{S_1}\right]^{-1}\mathbf{r}$$

$$=\left[(\mathbf{W}+\mathbf{S_2})\mathbf{W}^{-1}\mathbf{S_2}(\mathbf{W}+\mathbf{S_2})^{-1}\mathbf{W}+\mathbf{S_1}\right]^{-1}\mathbf{r}$$

$$=\left[(\mathbf{I}+\mathbf{S_2}\mathbf{W}^{-1})\mathbf{S_2}(\mathbf{I}+\mathbf{W}^{-1}\mathbf{S_2})^{-1}+\mathbf{S_1}\right]^{-1}\mathbf{r}$$

$$=\left[(\mathbf{S_2}+\mathbf{S_2}\mathbf{W}^{-1}\mathbf{S_2})(\mathbf{I}+\mathbf{W}^{-1}\mathbf{S_2})^{-1}+\mathbf{S_1}\right]^{-1}\mathbf{r}$$

$$=\left[\mathbf{S_2}(\mathbf{I}+\mathbf{W}^{-1}\mathbf{S_2})(\mathbf{I}+\mathbf{W}^{-1}\mathbf{S_2})^{-1}+\mathbf{S_1}\right]^{-1}\mathbf{r}$$

$$=[\mathbf{S_2}+\mathbf{S_1}]^{-1}\mathbf{r}$$
$$=\mathbf{S}^{-1}\cdot\mathbf{r}$$

Until now, we have proved that, VTM is convergent for 2-partite resistor network when using level-one wire tearing.

For the more general cases, the resistor network is *k*-partite, and the wire tearing might be level-one or multilevel. A general proof was presented in [62].

ACKNOWLEDGMENT

Discussions with Hao Zhang, Yi Su, Bin Niu, Yu Wang, Wei Xue, Peng Zhang and Chun Xia were very helpful. Thanks are due to Qi Wei, Bo Zhao, Xia Wei, Xiaojian Mao, Yongpan Liu, Fei Qiao and Rong Luo for encouragement and support. We are very grateful for the public domain softwares provided by John R. Gilbert, Tim A. Davis, Robert Bridson, Haifeng Qian and et al.

REFERENCES

[1] S. Balay, K. Buschelman, V. Eijkhout, W. D. Gropp, D. Kaushik, M. G. Knepley, L. Curfman McInnes, B. F. Smith and H. Zhang, PETSc Users Manual, ANL-95/11 - Revision 2.1.5, Argonne National Laboratory.

[2] R. Barrett, M. Berry, T. Chan, J. Demmel, J. Donato, J. Dongarra, V. Eijkhout, R. Pozo, C. Romine and H. Van der Vorst. Templates for the solution of Linear Systems: Building Blocks for Iterative Methods, 2nd Edition, SIAM, 1994.

[3] A. Basermann, U. Jaekel, and K. Hachiy. Preconditioning parallel sparse iterative solvers for circuit simulation. In Proceedings of the 8th SIAM Proceedings on Applied Linear Algebra, Williamsburg VA, 2003.

[4] A. Basermann, U. Jaekel, and M. Nordhausen. Parallel iterative solvers for sparse linear systems in circuit simulation. Fut. Gen. Fut. Gen.. Comput. Sys., 21(8):1275–1284, 2005.

[5] M. Benzi. Preconditioning techniques for large linear systems: A survey. Journal of Computational Physics, 182:418–477, 2002.

[6] J. Bolz, I. Farmer, E. Grinspun, and P. Schröoder. Sparse matrix solvers on the GPU: conjugate gradients and multigrid. In ACM SIGGRAPH 2003 (San Diego, California, July 27 - 31, 2003).

[7] C. W. Bomhof and H. A. Van der Vorst, A parallel linear system solver for circuit simulation problems, Numerical Linear Algebra with Applications, 7 (2000).

[8] F. H. Branin. Transient analysis of lossless transmission lines, Proceedings of the IEEE, vol. 55, pp. 2012–2013, Nov. 1967.

[9] R. Bridson. A MATLAB CMEX interface to the Metis library. Available at http://www.stanford.edu/~rbridson/download/metismex.c

[10] R. Bridson and W.-P. Tang. Refining an approximate inverse. Journal on Computational and Applied Math, 123 (2000), Numerical Analysis 2000 vol. III: Linear Algebra, pp. 293-306.

[11] S. Cauley, V. Balakrishnan, Cheng-Kok Koh, A parallel direct solver for the simulation of large-scale power/ground networks, IEEE Transactions on Computer-Aided Design of Integrated Circuits and Systems, v.29 n.4, p.636-641, April 2010.

[12] T. H. Chen and C. C.-P Chen. Efficient large-scale power grid analysis based on preconditioned Krylov-subspace iterative methods. In Proc. IEEE/ACM DAC, pages 559--562, 2001.

[13] R. E. Collin. Foundations for microwave engineering, 2nd edition, Wiley-IEEE Press, 2000.

[14] T. A. Davis and Y. F. Hu. The University of Florida Sparse Matrix Collection. Submitted to ACM Transactions on Mathematical Software. Available at: http://www.cise.ufl.edu/research/sparse/matrices/

[15] J. Demmel, J. Gilbert, and X. Li. An asynchronous parallel supernodal algorithm for sparse gaussian elimination. SIAM J. Matrix Analysis and Applications, 20(4): 915–952, 1999.

[16] Zhuo Feng, Peng Li. Multigrid on GPU: tackling power grid analysis on parallel SIMT platforms, Proceedings of the 2008 IEEE/ACM International Conference on Computer-Aided Design, November 10-13, 2008, San Jose, California.

[17] Zhuo Feng, Zhiyu Zeng. Parallel multigrid preconditioning on graphics processing units (GPUs) for robust power grid analysis. In Proceedings of the 47th Design Automation Conference (Anaheim, California, June 13 - 18, 2010). DAC '10. ACM, New York, NY, 661-666.

[18] T. L. Floyd. Principles of electric circuits, 6th edition, Prentice Hall, 1999.

[19] David Fritzsche, Andreas Frommer, and Daniel B. Szyld, Overlapping blocks by growing a partition with applications to preconditioning, Research Report 10-07-26, Department of Mathematics, Temple University, July 2010.

[20] N. Frohlich, B. M. Riess, U. A. Wever, and Q. Zheng A New Approach for Parallel Simulation of VLSI Circuits on a Transsitor Level. TCAD, June 1998.

[21] M. J. Gander, L. Halpern, and F. Nataf. Optimized Schwarz Methods. In T. Chan, T. Kako, H. Kawarada, O. Pironneau (eds.), Proceedings of the Twelveth International Conference on Domain Decomposition, DDM press, 2001, pp. 15–27.

[22] M. J. Gander, Schwarz Methods in the Course of Time, Electronic Transactions on Numerical Analysis, 31:228–255, 2008.

[23] A. George and J. W. Liu. Computer Solution of Large Sparse Positive Definite Systems. Prentice-Hall, Englewood Cliffs, New Jersey, 1981.

[24] John R. Gilbert, Gary L. Miller, and Shang-Hua Teng. Geometric mesh partitioning: Implementation and experiments. SIAM J. Scientific Computing 19:2091-2110, 1998.

[25] G. H. Golub and C. F. Van Loan, Matrix computations. Johns Hopkins University Press, 1989.

[26] Wei Huang. HotSpot: A chip and package compact thermal modeling methodology for VLSI design. Ph.D. Dissertation, University of Virginia, 2007.

[27] Russell Kao. Piecewise Linear Models for Switch-Level Simulation. Chapter 5.6.1, Node and Branch Tearing. Technical Report, CSL-TR-92-532, Stanford University, 1992.

[28] G. Karypis and V. Kumar. Metis 4.0: Unstructured graph partitioning and sparse matrix ordering system, Technical report, Department of Computer Science, University of Minnesota, 1998. Available at http://www.cs.umn.edu/metis

[29] D. P. Koester. Parallel Block-Diagonal-Bordered Sparse Linear Solvers for Power Systems Applications. Ph.D dissertation, Syracuse University. 1995.

[30] J. N. Kozhaya and S. R. Nassif. Fast Power Grid Simulation. In Proceedings of the 37th Design Automation Conference, 2000.

[31] J. N. Kozhaya, S. R. Nassif, and F. N. Najm, A multigrid-like technique for power grid analysis, IEEE Trans. Computer-aided Design, vol. 21, pp. 1148--1160, Oct. 2002.

[32] Zhao Li, C.-J. Richard Shi. A coupled iterative/direct method for efficient time-domain simulation of nonlinear circuits with power/ground networks. ISCAS (5) 2004: 165-168.

[33] Xiaoye S. Li. An overview of SuperLU: Algorithms, implementation, and user interface. ACM Trans. Math. Softw. 2005.

[34] Peng Li. What Is Parallel Circuit Simulation? ACM/SIGDA E-NEWSLETTER, vol. 40, No. 4, April 1, 2010

[35] P. L. Lions, On the Schwarz alternating method III: a variant for nonoverlapping subdomains, Third International Symposium on Domain Decomposition Methods for Partial Differential Equations, 1989, Houston, Texas.

[36] Sébastien Loisel and Daniel B. Szyld, On the convergence of Optimized Schwarz Methods by way of Matrix Analysis , Domain Decomposition Methods in Science and Engineering XVIII, Michel Bercovier, Martin Gander, Ralf Kornhuber, and Olof B. Widlund, editors. Lecture Notes in Computational Science and Engineering, Vol. 70, Springer, 2009, pages 363-370.

> REPLACE THIS LINE WITH YOUR PAPER IDENTIFICATION NUMBER (DOUBLE-CLICK HERE TO EDIT) <   14

[37] J. D. Meindl. Beyond Moore's Law: The Interconnect Era. Computing in Science and Engineering. 2003.

[38] L. Nagel. SPICE2: A Computer Program to Simulate Semiconductor Circuits, Electronics Research Laboratory Report No. ERL-M520. University of California, Berkeley, 1975.

[39] Jan Ogrodzki. Circuit simulation methods and algorithms. CRC Press, 1994.

[40] H. J. Pain. The physics of vibrations and waves, Wiley, 1976.

[41] He Peng, Chung-Kuan Cheng. Parallel transistor level full-chip circuit simulation. DATE 2009: 304-307.

[42] H. Qian, S. R. Nassif, and S. S. Sapatnekar, Power grid analysis using random walks, IEEE Trans. Computer-aided Design, vol. 24, pp. 1204--1224, Aug. 2005.

[43] H. Qian and S. S. Sapatnekar, Stochastic Preconditioning for Diagonally Dominant Matrices, SIAM Journal on Scientific Computing, Vol. 30, No. 3, pp. 1178 – 1204, March, 2008.

[44] T. L. Quarles. Analysis of Performance and Convergence Issues for Circuit Simulation. ERL Memo No. UCB/ERL M89/42 April 1989.

[45] Y. Saad. Iterative Methods for Sparse Linear Systems. The PWS Publishing Company, Boston, 1996. Second edition, SIAM, Philadelphia, 2003.

[46] R. A. Saleh, K. A. Gallivan, M. C. Chang, et al. Parallel circuit simulation on supercomputers. Proceedings of the IEEE, 1989.

[47] Alberto Sangiovanni-Vincentelli, Li-Kuan Chen, and Leon O. Chua. A new tearing approach – node-tearing nodal analysis. In IEEE International Symposium on Circuits and Systems, 1977.

[48] Jin Shi, Yici Cai, Wenting Hou, Liwei Ma, Sheldon X.-D. Tan, Pei-Hsin Ho, Xiaoyi Wang. GPU friendly fast Poisson solver for structured power grid network analysis, Proceedings of the 46th Annual Design Automation Conference, July 26-31, 2009, San Francisco, California.

[49] Ken Stanley, T. A. Davis. KLU: a "Clark Kent" sparse LU factorization algorithm for circuit matrices. 2004 SIAM Conference on Parallel Processing for Scientific Computing (PP04). Originally appeared in NA Digest, 1997.

[50] Heidi K. Thornquist, Eric R. Keiter, Robert J. Hoekstra, David M. Day, Erik G. Boman, A parallel preconditioning strategy for efficient transistor-level circuit simulation, Proceedings of the 2009 International Conference on Computer-Aided Design, November 02-05, 2009, San Jose, California.

[51] John R. Gilbert. Meshpart, a public domain matlab toolbox for sparse matrix partitioning. Available at
http://www.cerfacs.fr/algor/Softs/MESHPART/

[52] CircuitSim90, 1990 Circuit Simulation and Modeling Workshop at MCNC, Available at http://www.cbl.ncsu.edu/CBL_Docs/csim90.html

[53] Mathworks Corparation. Matlab User Manual. R13. 2002.

[54] Mathworks Corparation. Simulink User Manual. R13. 2002.

[55] A. Toselli and O. Widlund. Domain Decomposition Methods – Algorithms and Theory. Springer Series in Computational, Mathematics 34, Springer, Berlin, Heidelberg, 2005.

[56] R. S. Varga. Matrix Iterative Analysis. Prentice-Hall, Englewood Cliffs, New Jersey, 1962. Second Edition, Springer Series in Computational Mathematics 27, Springer, Berlin, Heidelberg, New York, 2000.

[57] W. T. Weeks, A. J. Jimenez, G. W. Mahoney, and D. Mehta. Algorithms for ASTAP - A Network-Analysis Program. IEEE Trans.Circuit Theory, CT-20(4):628-634, 1973.

[58] Fei Wei, Huazhong Yang. Directed Transmission Method, A Fully Asynchronous Approach to Solve Sparse Linear Systems in Parallel. In ACM Proceedings of the 20th Symposium on Parallelism in Algorithms and Architectures (Munich, Germany, June 14 - 16, 2008). SPAA 2008.

[59] Fei Wei, Huazhong Yang, Virtual Transmission Method, A New Distributed Algorithm to Solve Sparse Linear Systems. The Fourth International Conference on Networked Computing and Advanced Information Management, vol. 1, pp.703-709, 2008.

[60] Fei Wei, Huazhong Yang. Waveform Transmission Method, A New Waveform Relaxation Based Algorithm to Solve Ordinary Differential Equations. Preprinted. 2009.

[61] Fei Wei, Huazhong Yang. Transmission Line Inspires a New Distributed Algorithm to Solve the Nonlinear Dynamical System of Physical Circuits. The 5th International Conference on Computer Sciences and Convergence Information Technology, Nov. 30, Seoul, Korea, 2010.

[62] Fei Wei, Huazhong Yang. Virtual Transmission Method, Algorithm and Theory. Preprinted. 2009.

[63] J. White, A. Sangiovanni-Vincentelli, Relaxation Techniques for the simulation of VLSI circuits. Kluwer Academic Publishers, 1986.

[64] Felix F. Wu. Solution of large-scale networks by tearing. IEEE Transactions on Circuits and Systems, 1976.

[65] Wei Xue, Jiwu Shu, Yongwei Wu, Weimin Zheng. Parallel Algorithm and Implementation for Realtime Dynamical Simulation of Power System. ICPP 2005: 137-144.

[66] D. M. Young. Iterative Solution of Large Linear Systems. Academic Press, New York, 1971.

[67] Yu Zhong, M. D. F. Wong, Fast algorithms for IR drop analysis in large power grid, Proceedings of the 2005 IEEE/ACM International conference on Computer-aided design, p.351-357, November 06-10, 2005.

[68] Quming Zhou, Kai Sun, Kartik Mohanram, Danny C. Sorensen. Large power grid analysis using domain decomposition, Proceedings of the conference on Design, automation and test in Europe: Proceedings, March 06-10, 2006, Munich, Germany.

[69] C. Zhuo, J. Hu, M. Zhao, and K. Chen. Power grid analysis and optimization using algebraic multigrid. IEEE Transactions on Computer-aided Design, 27:738–751, April 2008.

**Fei Wei** received the B.S. degree in electronic engineering from Tsinghua University, Beijing, China in 2004. He is currently a Ph.D student in Department of Electronic Engineering, Tsinghua University. His research interests are distributed numerical algorithms and transistor-level circuit simulation.

**Huazhong Yang** (M'97–SM'00) received the B.S. degree in microelectronics and the M.S. and Ph.D. degrees in circuits and systems from Tsinghua University, Beijing, China, in 1989, 1993, and 1998, respectively. Since 1993, he has been with the Department of Electronic Engineering, Tsinghua University, where he has been a Full Professor since 1998. His research interests include CMOS radio-frequency integrated circuits, VLSI system structure for digital communications and media processing, low-voltage and low power integrated circuits, and computer-aided design methodologies for system integration. He has authored or coauthored 6 books and more than 180 journal and conference papers in this area and holds 9 China patents. He is also a coeditor of the research monograph High-speed Optical Transceivers-Integrated Circuits Designs and Optical Devices Techniques (World Scientific, 2006).

Dr. Yang was a recipient of the fund for Distinguished Young Scholars from NSFC in 2000, the outstanding researcher award of the National Keystone Basic Research Program of China in 2004, and the Special Government Allowance from the State Council of China in 2006.served as a TPC member of the Asia-Pacific Conference on Circuits and Systems, the International Conference on Communications, Circuits and Systems, and the Asia and South Pacific Design Automation Conference. He is an Associate Editor of the International Journal of Electronics.